\documentclass[pra,twocolumn,showpacs,preprintnumbers]{revtex4}
\usepackage{graphicx}
\usepackage{amsfonts}
\usepackage{amssymb}
\usepackage{amsmath}
\usepackage{bm}
\usepackage{subfigure}

\newcommand{\be}{\begin{equation}}
\newcommand{\ee}{\end{equation}}
\newcommand{\bea}{\begin{eqnarray}}
\newcommand{\eea}{\end{eqnarray}}
\newcommand{\Eq}[1]{Eq.\,(\ref{#1})}% \Eq{abc}
\newcommand{\Fig}[1]{Fig.\,\ref{#1}}% \Fig{fig:abc}
\newcommand{\Sec}[1]{Sec.\,\ref{#1}}% \Sec{sec:abc} sic!byc konsekewntnym \label{sec:xx} \Sec{sec:xx}
\begin{document}

\title{Resonant state expansion applied to one-dimensional quantum systems}

\author{ A.\,Tanimu}
\author{E.\,A. Muljarov}
\affiliation{School of Physics and Astronomy, Cardiff University, Cardiff CF24 3AA, United Kingdom}
\begin{abstract}
The resonant state expansion, a rigorous perturbation theory,  recently developed in electrodynamics, is applied to non-relativistic quantum mechanical systems in one dimension.  The method is used here for finding the resonant states in various potentials approximated by combinations of Dirac delta functions. The resonant state expansion is first verified for a triple quantum well system, showing convergence to the available analytic solution as the number of resonant states in the basis increases. The method is then applied to multiple quantum well and barrier structures, including finite periodic systems. Results are compared with the eigenstates in triple quantum wells and infinite periodic potentials, revealing the nature of the resonant states in the studied systems.
\end{abstract}


\date{\today}

\maketitle

\section{Introduction}

The resonant state expansion (RSE) is a rigorous perturbative method for treating open optical systems, which has been recently developed in electrodynamics~\cite{Egor}. The RSE has been verified and applied to various one-dimensional (1D), 2D and 3D open optical systems~\cite{Doost4,Doost5,Armitage14,Doost6,EMuljarov8}, demonstrating high efficiency of the method and its suitability for treating perturbations of the permittivity of arbitrary strength and shape. The RSE treats the perturbed problem as a combination of an unperturbed one, usually having an analytical solution, and a perturbation. It is well known that the existence of a continuum of states in the spectrum of a system presents a significant problem for any perturbation theory. In open quantum systems such a continuum is often the dominating if not the only part of the spectrum. However, going away from the real axis to the complex frequency plane, the continuum can in many cases be effectively replaced by a countable number of discrete resonant states (RSs). In optics, these are vectorial eigen solutions of Maxwell's equations. In non-relativistic quantum mechanics, where the concept of RSs was originally introduced in the pioneering works of Gamow~\cite{Gamow} and Siegert~\cite{Siegert}, RSs are described by complex scalar wave functions.
%The concept of RSs, however, originates from non-relativistic quantum mechanics, dealing with scalar wave functions. It was introduced in the original works of Gamow~\cite{Gamow} and Siegert~\cite{Siegert}, see also review articles~\cite{Garcia,Moiseyev}.

Quantum-mechanical RSs are the eigen solutions of the Schr\"{o}dinger wave equation with purely outgoing wave boundary conditions~\cite{Siegert,Gamow,Garcia,Moiseyev}.
RSs have generally complex energy eigenvalues $E_n=\hbar \Omega_n-i\hbar\Gamma_n$, with a negative imaginary part having the meaning of the inverse lifetime of the quantum state, for which the wave function  is exponentially decaying in time as  $e^{-\Gamma_n t}$. Already in the early works on RSs, it has been understood that owing to this decay, the wave function grows in space exponentially at large distances, reflecting the fact that the probability density leaks out of the open system~\cite{Zel'dovich2}. These exponentially increasing tails of RSs outside the system make the wave function not square integrable, thus preventing from using the standard normalization condition. Therefore, a special normalization of RSs was proposed~\cite{Siegert,Snoll,Garcia}. With this normalization, the RSs can then be used to calculate the Green's function of the system via its spectral representation based on the Mittag-Leffler theorem\cite{Newton,More}. The Green's function in turn fully describes the linear response of the system and allows to calculate its observables, such as the local density of states, scattering, and transmission.

It has been also realized~\cite{Newton,More} that the full set of the RSs is complete within the finite area of space occupied by an open system, and therefore the RSs can be used as a basis for expanding solutions of the  Schr\"{o}dinger equation, also with modified potentials.  Using this approach, the Schr\"{o}dinger wave equation is reduced to a linear matrix eigenvlue problem, which can be solved by diagonalizing a complex symmetric matrix. This 1D quantum-mechanical analog of the RSE was formulated in~\cite{Bang} with the only numerical implementation known in the literature~\cite{PLind}, which was a calculation of a single bound state in a rectangular quantum well. The conclusion made in~\cite{PLind} was that the convergence of this approach is not sufficiently quick compared to other methods, also considered in~\cite{PLind}. Perhaps, this message has become one of the reasons why this approach was not developed any further in quantum mechanics.

Very recently, the RSE has been independently re-invented in electromagnetics~\cite{Egor}, with several significant differences compared to the original concept~\cite{Bang}, which are mainly related to the vectorial nature of the electromagnetic field and relativistic form of the Maxwell wave equation~\cite{Egor,Doost4,Doost5,Armitage14,Doost6,EMuljarov8}. It has also been shown~\cite{Doost6,Lobanov} that the RSE actually presents a very efficient computational tool, with a potential to supersede some popular computational methods, such as finite-difference time-domain, finite element method, Fourier modal method, an so on. This indicates clearly that there is a need also to study the quantum-mechanical (QM) analogue of the RSE (QM-RSE), which we start doing in the present work.

The aim of this paper is to apply, verify and study the QM-RSE in various simple 1D quantum systems. To facilitate the analytics, we have employed the model of Dirac delta functions for describing quantum potentials. We first calculate the RSs of a symmetric double quantum well structure modeled by delta functions. These RSs are then taken as an unperturbed basis for the QM-RSE. Both symmetric and asymmetric triple quantum well or barrier structures, which allow relatively simple analytic solutions, are used to verify the QM-RSE and to study its convergence. In this case, the potential of the third well or barrier in the middle is treated as a perturbation. The QM-RSE is then used for calculation of the RSs in multiple quantum well structures and finite periodic quantum lattices of different period and potential strength.

%{\bf EM: Use Bibtex for references.}

\section{Formalism of the QM-RSE}
\label{FRSE}
In this section we outline the formalism of the QM-RSE which has been developed earlier in~\cite{Bang} and \cite{Egor}. The QM-RSE treats a perturbation $\Delta V(x)$ of the quantum potential in the time-independent Schr\"{o}dinger equation,
\be
\left[-\frac{\hbar^2}{2m }\frac{d^2}{dx^2}+V(x)+\Delta V(x)\right]\psi_\nu(x)=E_\nu \psi_\nu(x)\,,
\label{pertSE}
\ee
by using as a basis the RSs for the unperturbed potential $V(x)$ and transforming \Eq{pertSE} into a matrix eigenvalue problem.
Here, for simplicity, we concentrate on the 1D Schr\"{o}dinger equation for a particle with mass $m$. $\psi_\nu(x)$ and $E_\nu$  are, respectively, the wave functions and the energies of the perturbed RSs, and the index $\nu$ is used to label different RSs of the particle. It is useful to introduce also the RS wave numbers $\varkappa_\nu$ defined as  $E_\nu=\hbar^2\varkappa_\nu^2/(2m)$. The corresponding wave functions and the wave numbers of the RSs in the unperturbed potential $V(x)$ are denoted by $\varphi_n(x)$ and $k_n$, respectively, where the index $n$ labels the unperturbed RSs. In full analogy with the RSE in optics~\cite{Egor,Doost6} where the permittivity would play the role of the quantum potential in the wave equation, the QM-RSE is applicable to potentials with compact support and perturbations included in the area occupied by the unperturbed system.

Using the Green's function of the Schr\"{o}dinger equation for the unperturbed quantum potential $V(x)$ and treating the term with perturbation $\Delta V(x)\psi_\nu(x)$ in \Eq{pertSE} as an inhomogeneity, one can find a formal solution of \Eq{pertSE}. Then, applying the Mittag-Leffler expansion to the Green's function, and expanding the perturbed RSs into the unperturbed ones,
\be
\psi_{\nu}(x)=\sum_n C_{n\nu} \sqrt\frac{\varkappa_\nu}{k_n}\varphi_n(x)\,,
\label{pertwave}
\ee
the Schr\"{o}dinger equation (\ref{pertSE}) is converted into a linear complex eigenvalue problem~\cite{Bang,Egor}
\be
\sum_m H_{nm}C_{m\nu}=\varkappa_\nu C_{n\nu}\,,
\label{RSE}
\ee
where
\be
H_{nm}= k_n\delta_{nm}+\frac{\Delta V_{nm}}{2\sqrt{k_n}\sqrt{k_m}}\,,
\label{Hnm}
\ee
\be
 \Delta V_{nm}=\int_{-a}^{a}\varphi_n(x) \Delta V(x)\varphi_m(x)dx\,,
\label{Vnm}
\ee
and $\delta_{nm}$ is the Kronecher delta. Here we assumed, without loss of generality, that the perturbation is located within the region $|x|\leqslant a$.

The perturbed wave numbers $\varkappa_\nu$  and the expansion coefficients $C_{n \nu}$ can be found by diagonalizing the complex symmetric matrix $H_{nm}$, consisting of the diagonal matrix of the unperturbed eigen wave numbers $k_n$ and the perturbation matrix $\Delta V_{nm}$.  The $\sqrt{k_n}$ factors are introduced in Eqs.\,(\ref{pertwave}) and (\ref{Hnm}) in order to symmetrize the eigenvalue problem.

The perturbation matrix \Eq{Vnm} is determined by the unperturbed wave functions $\varphi_n(x) $ which have to be properly nomalized. As shown in~\cite{Bang,Egor,Doost6}, the proper normalization in 1D, leading to the eigenvalue problem \Eq{RSE}, has the following  form:
\be
1=\int_{-a}^{a}\varphi^2_n(x)dx-\frac{\varphi^2_n(a)+\varphi^2_n(-a)}{2ik_n}\,,
\label{gennorm}
\ee
where we have used the fact that the inhomogeneity of the unperturbed potential is located within the region $|x|\leqslant a$, so that $x=\pm a$ are the boundaries of the unperturbed open quantum system. It can be seen~\cite{Tanimu} that for bound states, \Eq{gennorm} is equivalent to the standard normalization  condition
$\int_{-\infty}^{\infty}\varphi^2_n(x)dx=1$, in which case the wave function of a bound state can always be taken real.
A more detailed discussion of the normalization of the RSs in quantum-mechanical systems can be found in~\cite{Snoll,More,More2,Watson}.

The complete basis of RSs usually contains an infinite countable number of functions. Therefore, the matrix equation (\ref{RSE}) of the QM-RSE has infinite size and for practical use requires a truncation. This truncation presents the only limitation of the QM-RSE, thus making it an asymptotically exact method. Moreover, as it was demonstrated in~\cite{Egor}, owing to its quick convergence, the RSE is capable of treating arbitrarily strong perturbations, provided that a sufficient number of RSs is kept in the basis, in order to guarantee the required accuracy of calculation.

\section{Unperturbed resonant states: Double quantum well}
\label{unptbed}

To apply the QM-RSE for particular quantum systems, we need to choose a suited basis of RSs. These are the solutions of the Schr\"{o}dinger equation with an unperturbed potential $V(x)$ which in principle can be chosen arbitrary, though both $V(x)$ and $\Delta V(x)$ have to be functions with compact support, and the perturbation $\Delta V(x)$ must be non-vanishing only within the area of inhomogeneity of $V(x)$,  as already noted. Usually, the optimal choice of the unperturbed potential is such that the Schr\"{o}dinger equation with $V(x)$ has an analytic solution and at the same time is close to the full potential to be treated, in this way minimizing the effect of the perturbation.

In this work, however, we have chosen as unperturbed, or the basis system the most simple 1D quantum potential containing RSs: a double symmetric quantum well described by two Dirac delta functions. We fix this choice for all perturbed examples considered below, varying only the parameters of the basis system, where necessary. We also use the convenient units of $\hbar=1$ and $m=1/2$ throughout this work.

The unperturbed quantum potential is thus given by
\be
V(x)=-\gamma\delta(x-a)-\gamma\delta(x+a)\,,
\label{unppot}
\ee
which models a symmetric double quantum well (barrier) structure
for $\gamma>0$ ($\gamma<0$). Here, $\delta(x)$ is the Dirac delta function,
$2a$ is the distance between the wells, and $\gamma$ is the strength of the potential which has the meaning  of the depth of each quantum well multiplied by its width, bearing in mind a comparison of this model with a corresponding pair of rectangular quantum wells. An obvious advantage of the model is its simplicity and explicit analytical solvability. The solution of the unperturbed Schr\"{o}dinger equation is given by~\cite{Tanimu}
\begin{equation}
\varphi_n(x)=
\begin{cases}
A_n e^{ik_nx},                               &  x> a,\\
B_n( e^{ik_nx}\pm e^{-ik_nx})         & |x| \leqslant a,\\
\pm A_n e^{-ik_nx},                               & x<-a,\\
\end{cases}
\label{SWE}
\end{equation}
where the basis RS wave numbers $k_n$ satisfy the secular equation
\be
1+ \frac{2ik_n}{\gamma}=\mp e^{2ik_na}\,,
\label{unptseq}
\ee
with the upper (lower) sign corresponding to even (odd) parity states. Note that \Eq{unptseq} generates a complete set of basis RSs which include bound, antibound and normal RSs, as classified and discussed in detail in~\cite{Tanimu}. They are all required for the completeness and thus have to be taken into account in the QM-RSE.

The normalization of RSs, which was also calculated in~\cite{Tanimu}, using the definition \Eq{gennorm}, is given by
\bea
A_n&=&B_n\left(1+\frac{\gamma}{2ik_n}\right)^{-1},
\\
B_n&=&\frac{1}{2\sqrt{\pm[a-(\gamma+2ik_n)^{-1}]}}\,.
\label{norm-double}
\eea

\section{Verification of the QM-RSE: Triple quantum wells }
\label{Sec:VerfRSE}

To verify the QM-RSE and to study its convergence, we take another exactly solvable system, having a relatively simple analytic solution: a triple quantum well described by three delta functions. For simplicity, we keep the strength of the left and right wells/barriers (separated by the distance $2a$) the same, whereas the position and the strength of the middle well/barrier can be any. In this way, our triple well/barier system is described by the potential $V(x)+\Delta V(x)$, where $V(x)$ is given by \Eq{unppot} and
\be
\Delta V(x)=-\beta\delta(x-b)\,,
\label{triplpot}
\ee
with $|b|<a$. As noted above, the position $b$ and the strength $\beta$ of the middle well/barrier are arbitrary parameters, with $\beta>0$ corresponding to a well and $\beta<0$ to a barrier.

\subsection{Analytic solution}
\label{ansol}
A general analytic solution of the Schr\"{o}dinger equation with the triple delta potential $V(x)+\Delta V(x)$, where $V(x)$ and $\Delta V(x)$ are given by Eqs.\,(\ref{unppot}) and (\ref{triplpot}), respectively, is also provided in~\cite{Tanimu}. The secular equation for the RS wave numbers $\varkappa$ of this perturbed quantum system is given by
\be
\xi^2 (1-\eta) - 2\xi \cos (2\varkappa b) +1+\eta=0\,,
\label{triple}
\ee
where
\be
\xi=\frac{e^{2i\varkappa a}}{1+2i\varkappa/\gamma}\,,\ \ \ \ \ \eta=\frac{2i\varkappa}{\beta}\,.
\label{xi1}
\ee

\begin{figure}[t]
    \centering
%\vskip-0.6cm %remove
\vskip0.3cm
\hskip-2.6cm
 \includegraphics[scale=0.35,angle=-90]{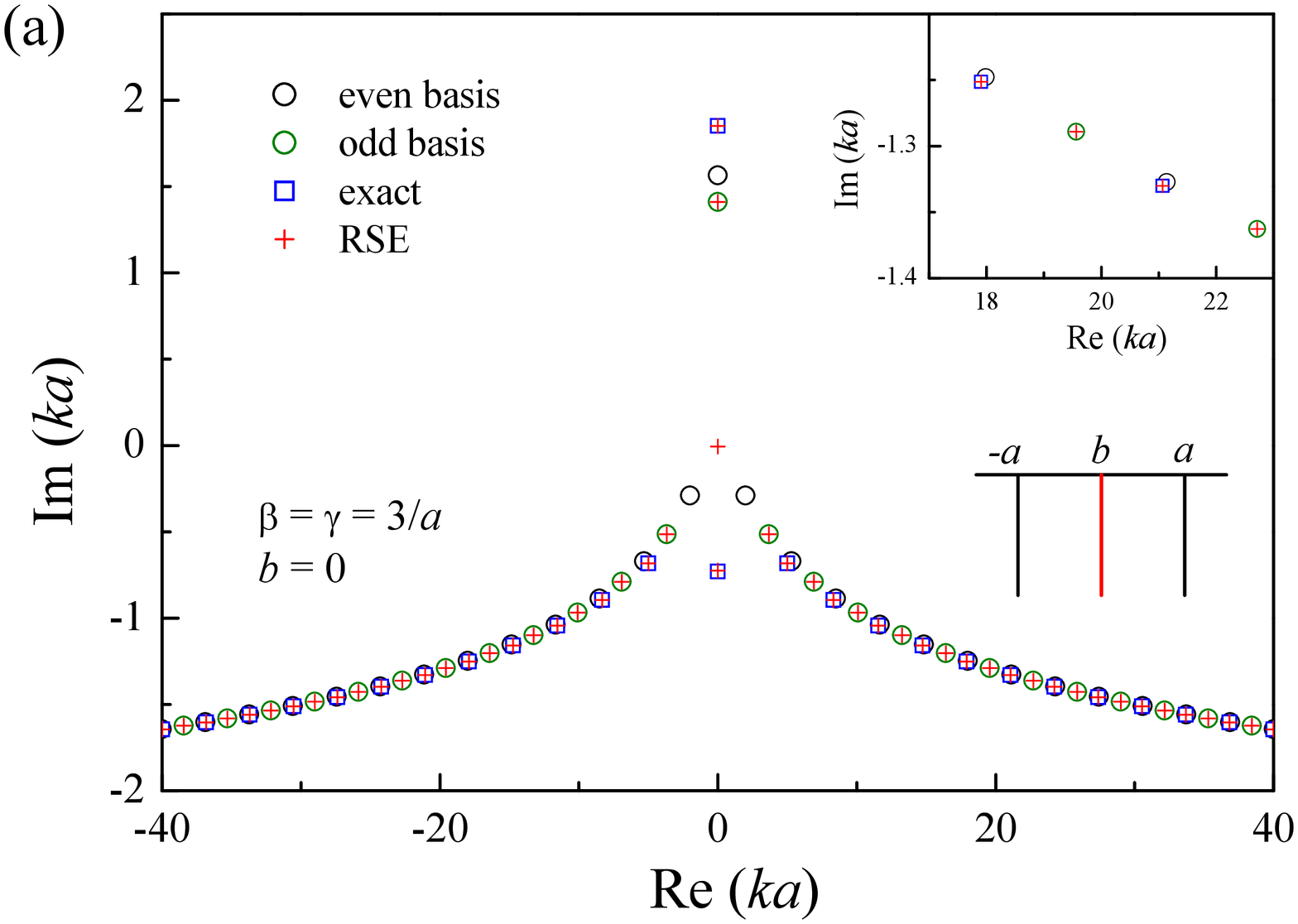}
\vskip-2.0cm
\hskip-2.6cm
 \includegraphics[scale=0.35,angle=-90]{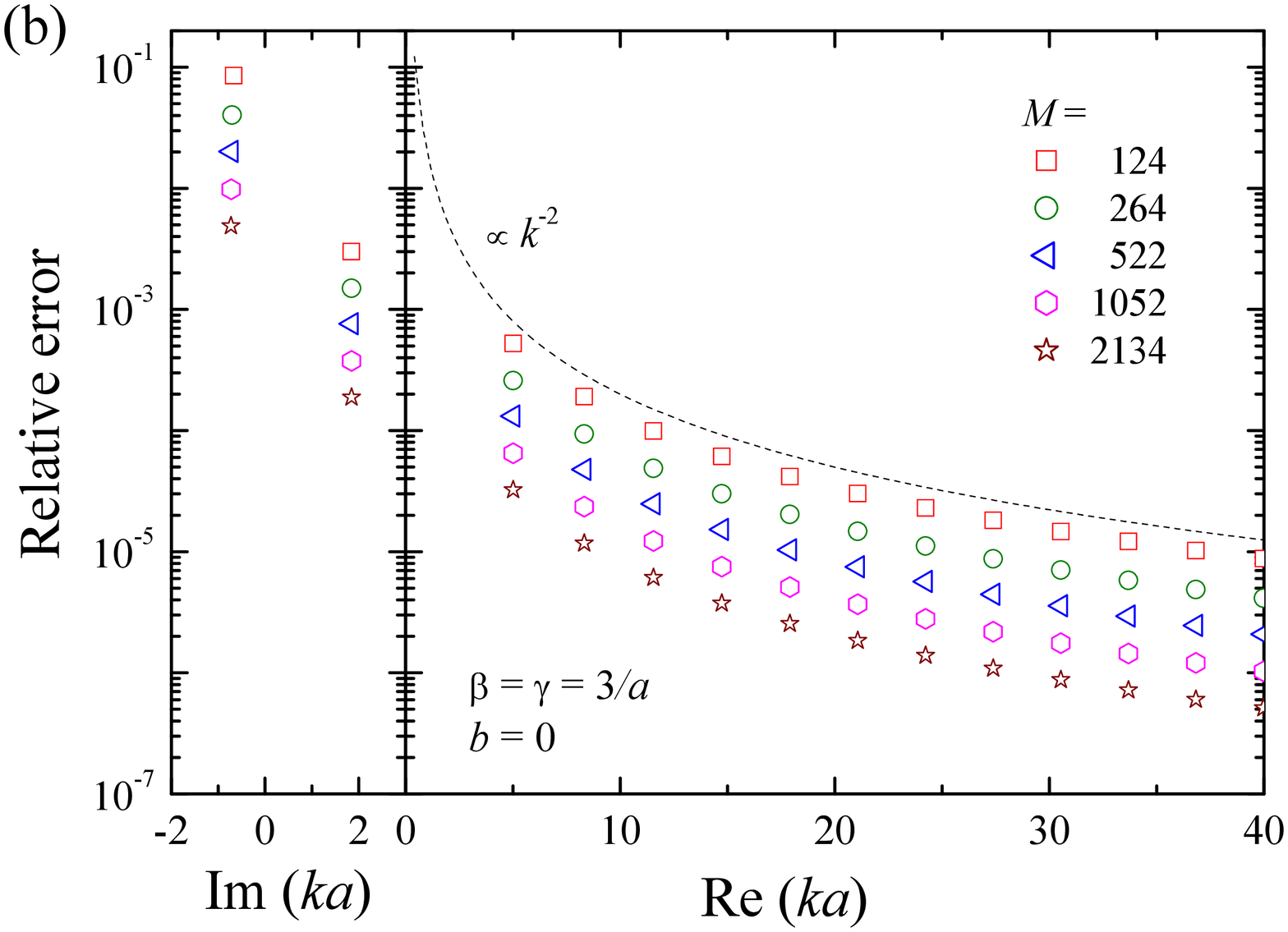}
\vskip-2.3cm
    \caption{(a) Eigen wave numbers of the resonant states for a double symmetric quantum well with $\gamma=3/a$ (open circles) and a triple symmetric quantum well with $b=0$ and $\gamma=\beta=3/a$, calculated using the QM-RSE (red crosses) and the analytic secular equation (\ref{triple}) (blue squares). The wave numbers of even (black circles) and odd (green circles) unperturbed RSs for a double quantum well structure are calculated via Eq.\,(\ref{unptseq}).  The insets show a zoom-in of a particular area and a sketch of the perturbed potential (black lines) and the perturbation used (red line).
    (b) Relative error of the QM-RSE values of the RS wave numbers as function of the real or imaginary part of the wave number, for different basis sizes $M$ as given. The dashed line shows a power law dependence as labeled.  }
\label{fig:fig1RSE}
\end{figure}

\begin{figure}[t]
    \centering
\vskip0.3cm
\hskip-2.6cm
 \includegraphics[scale=0.35,angle=-90]{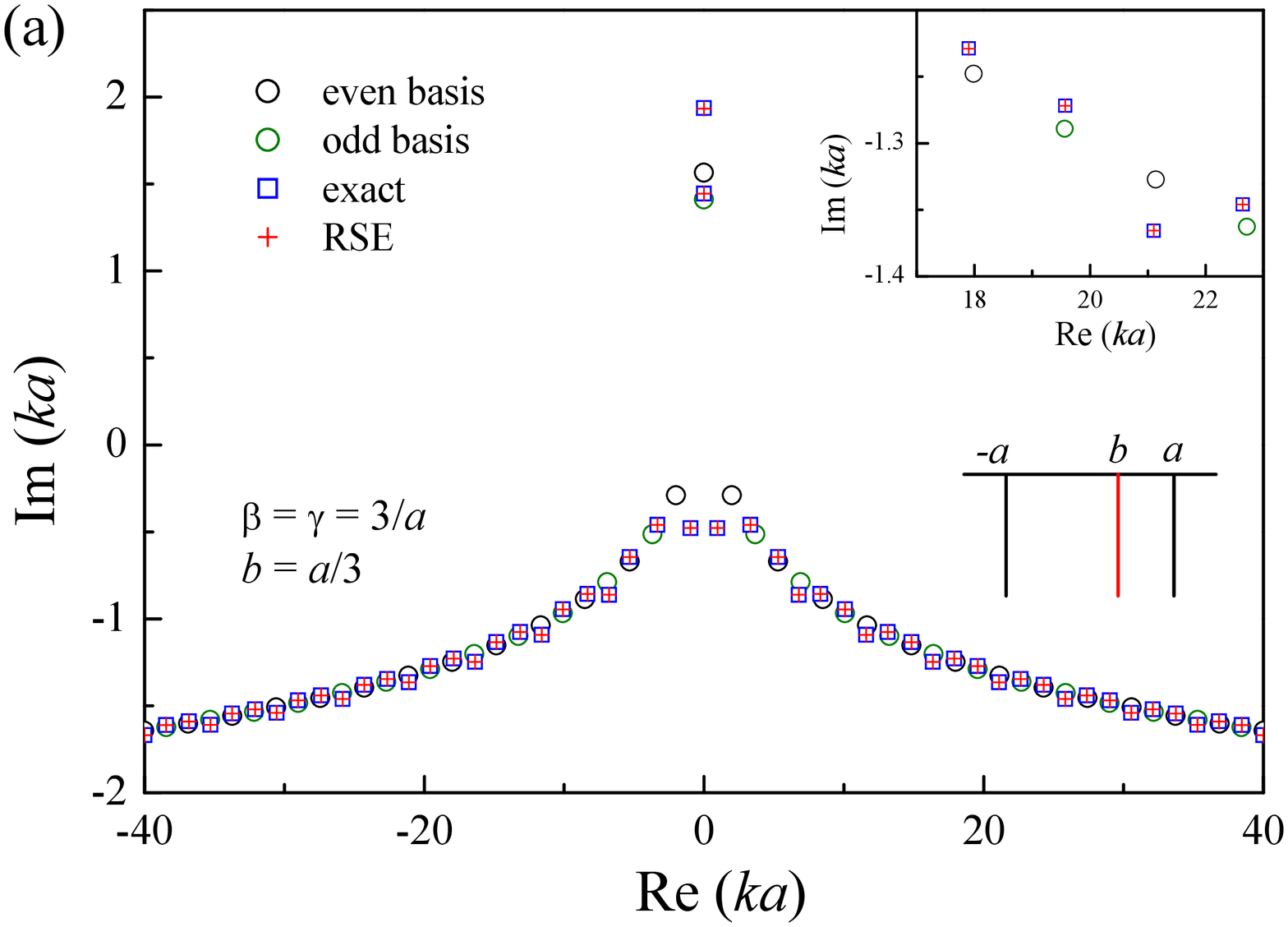}
\vskip-2.0cm
\hskip-2.6cm
 \includegraphics[scale=0.35,angle=-90]{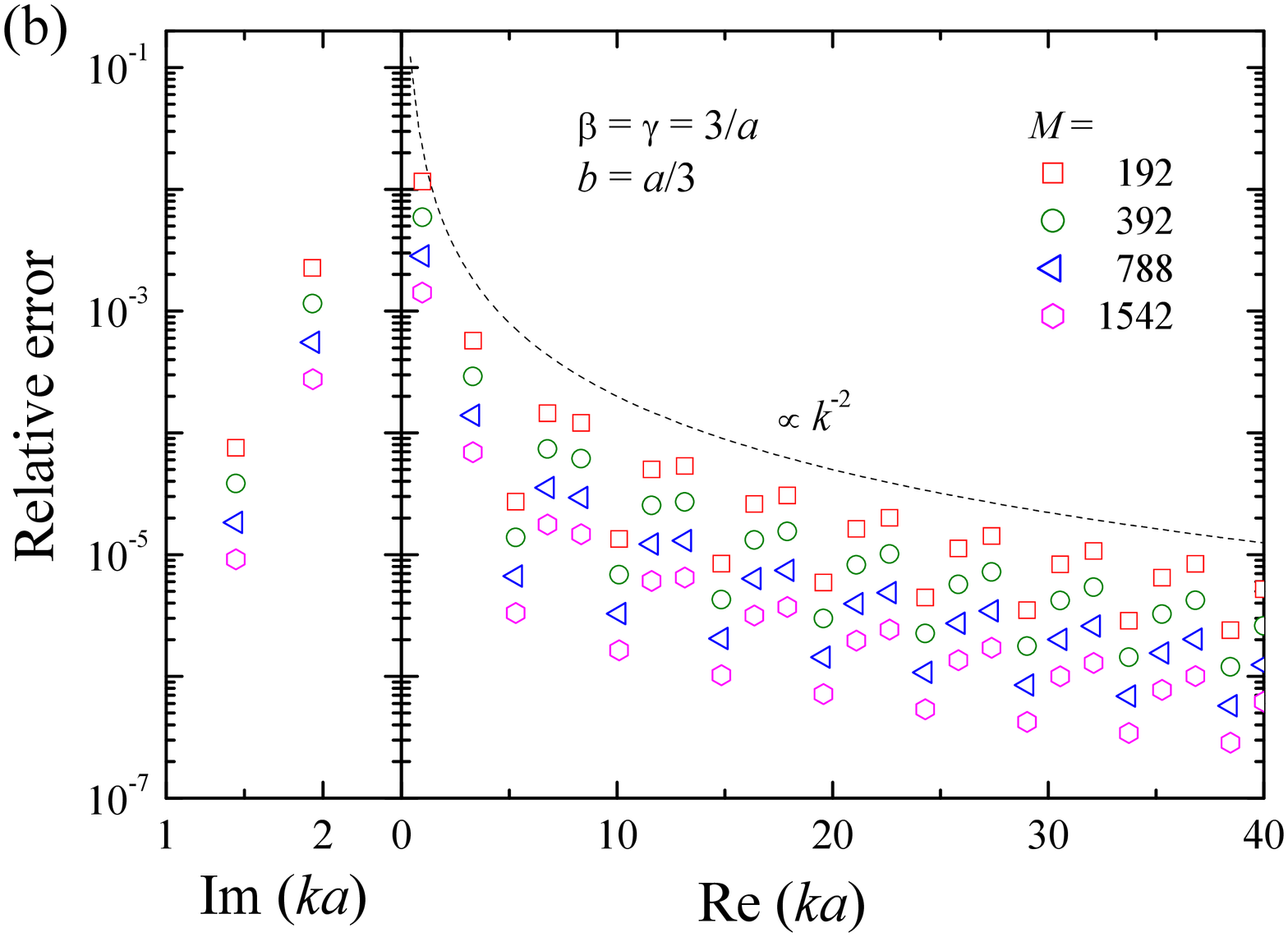}
\vskip-2.3cm
    \caption{As \Fig{fig:fig1RSE} but $b=a/3$.
    }
\label{fig:figxRSE}
\end{figure}

\begin{figure}[t]
    \centering
\vskip0.3cm
\hskip-2.6cm
 \includegraphics[scale=0.35,angle=-90]{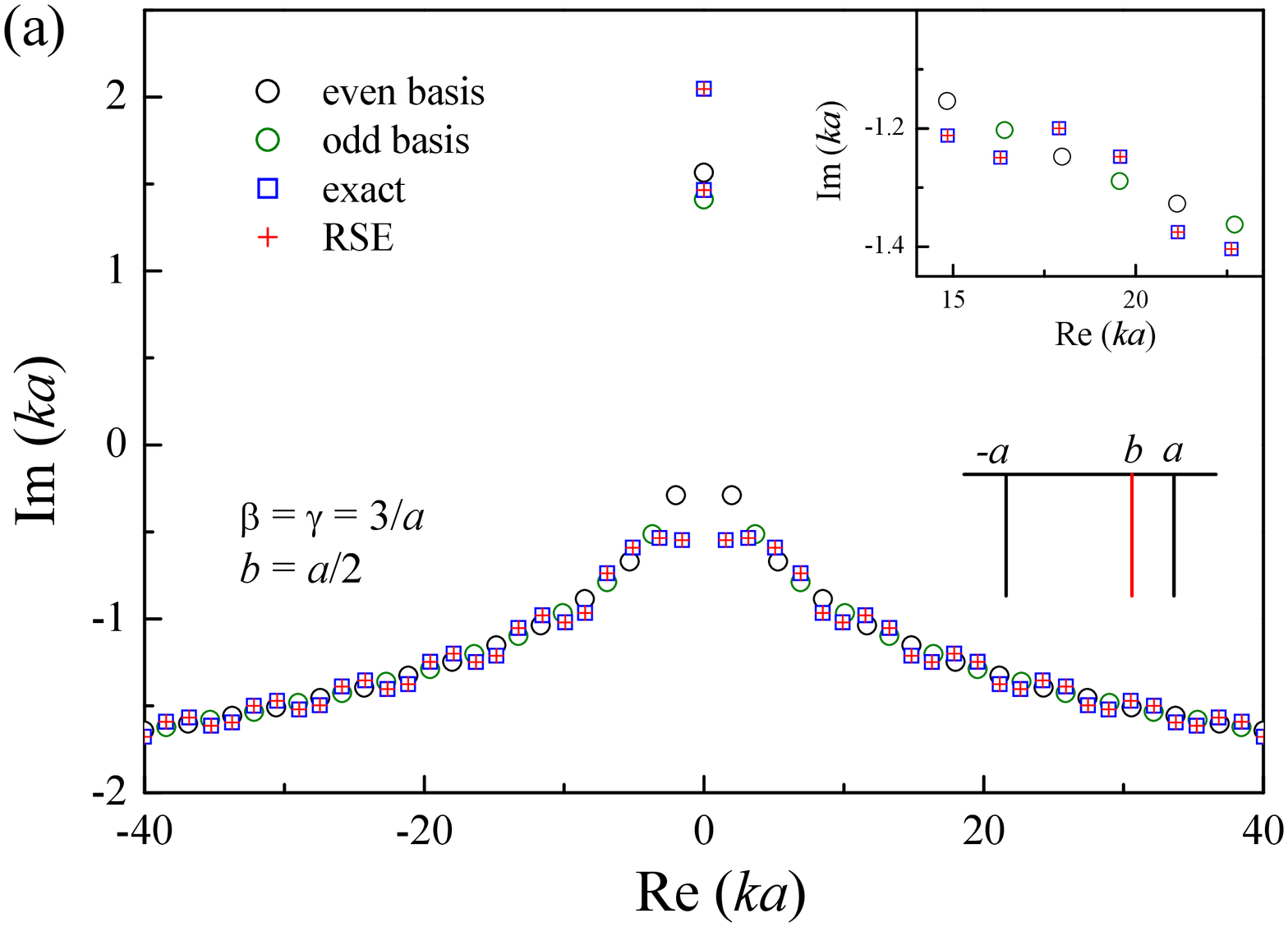}
\vskip-2.0cm
\hskip-2.6cm
 \includegraphics[scale=0.35,angle=-90]{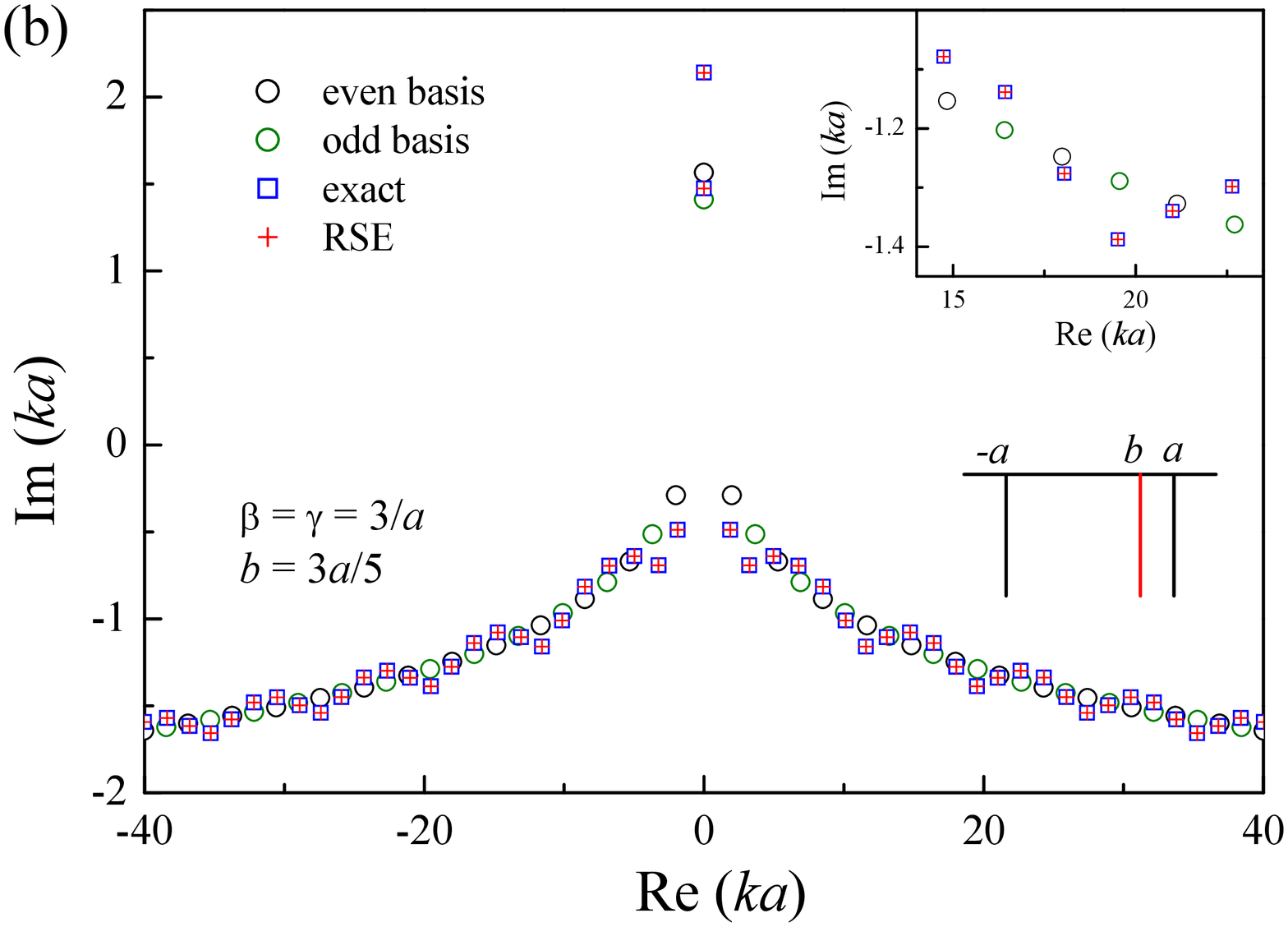}
\vskip-2.0cm
\hskip-2.6cm
 \includegraphics[scale=0.35,angle=-90]{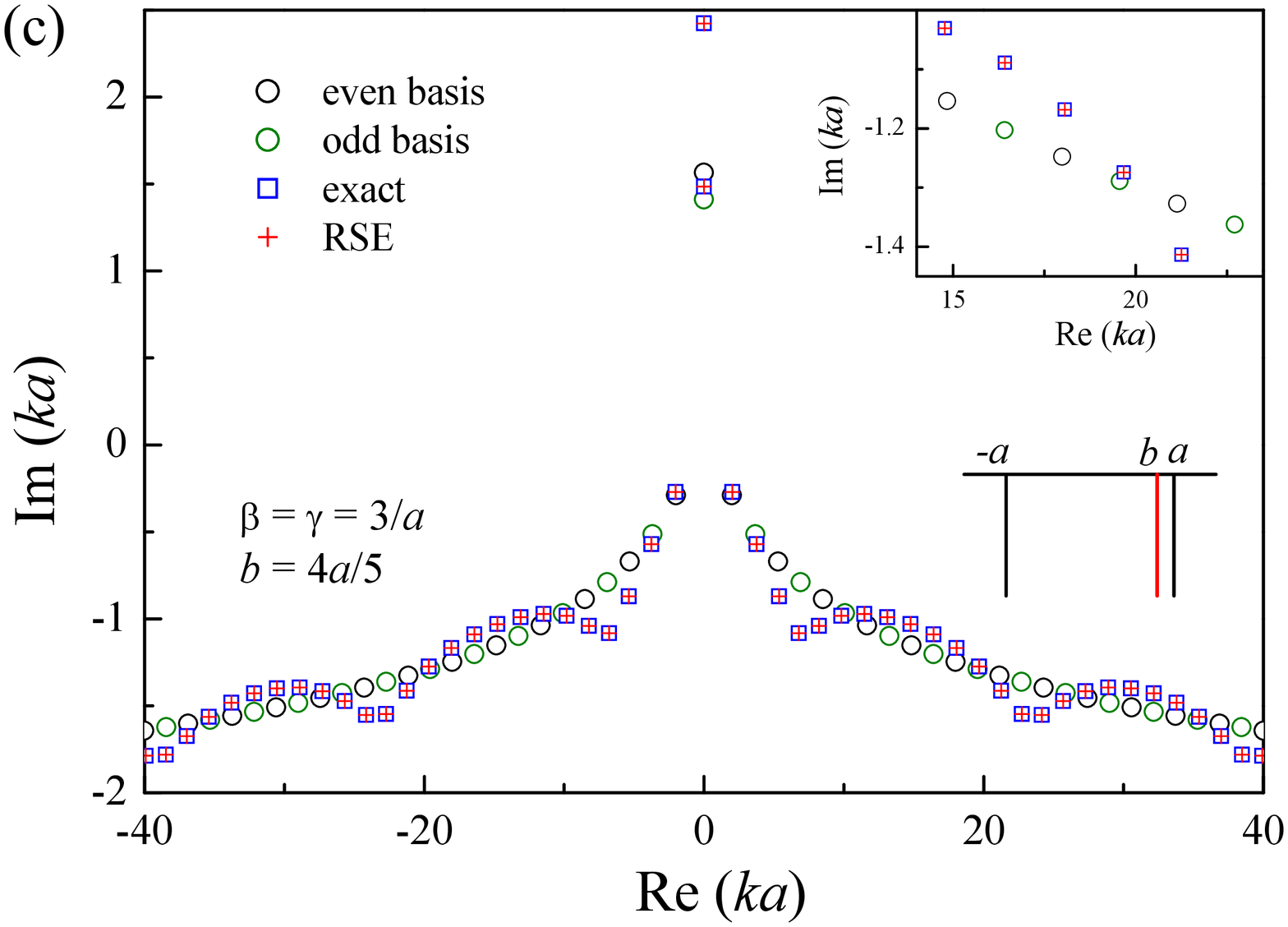}
\vskip-2.3cm
    \caption{As  \Fig{fig:figxRSE}\,(a) but for $b=a/2$ (a), $b=3a/5$ (b), and $b=4a/5$ (c).
}
\label{fig:fig2RSE}
\end{figure}

Both secular equations (\ref{unptseq}) and (\ref{triple}), for double and triple quantum wells, are solved numerically to find the exact RSs wave numbers of the unperturbed and perturbed problem, respectively. This is done using the Newton-Raphson method implement in MATLAB. Exact wave numbers of the RSs in double and triple quantum wells are presented in the complex $k$-plane in Figs.~\ref{fig:fig1RSE}--\ref{fig:fig2RSE} and compared with QM-RSE. Before looking at this comparison (which is discussed in \Sec{b0} and \Sec{asym} below), we would like to concentrate on the physical results.
%
%{\bf (i) The data in (a) is wrong. Probably it is for the barrier at the center. (ii) Which value of $\beta$ is used in reality? The changed in the spectrum seems too small for $\beta=10/a$. (iii) Is $M$ in (b) the total number of even states or the total number of even and odd states together?  (iv) Need to take RSs symmetrically on both sides, as described in the text. This will lead to either even or odd $M$. Please check and correct, providing precise numbers for M. (v) In the inset, make the sketch for the potential more realistic, keeping approximately the potential ratio which is 3 to 10.}

Since a single delta-function potential well always has only one bound state for any strength of the potential ($\gamma>0$), it is clear that a double-delta potential can accommodate two bound states at most~\cite{Tanimu}. For the parameters used for the double well ($\gamma=3/a$), these two bound states are present in the spectrum and are seen  in \Fig{fig:fig1RSE}\,(a) on the imaginary $k$-axis (black and green circles). All other eigenmodes of the double well system are the normal RSs which always exist in pairs, thus providing the mirror symmetry of the full spectrum of RSs in the complex $k$-plane, which is a general property of any open system. As it follows from the mathematical solution of Eqs.\,(\ref{unptseq}) and (\ref{triple}), there is an infinite countable number of RSs in the spectrum, which is another general property of an open system. Furthermore,  \Fig{fig:fig1RSE}\,(a) shows that the normal RSs are almost equally spaced when the real part of $k_n$ is much larger than the imaginary part. This quasi-periodicity of the RS wave numbers can be understood as a result of constructive interference of quantum waves propagating back and forth within the system and experiencing multiple reflections from the wells/barriers at the boundaries $x=\pm a$. From this resonant condition for the constructive interference one can estimate the distance between the neighboring RSs in the complex $k$-plane as $\sim\pi/(2a)$, which is observed in \Fig{fig:fig1RSE}\,(a).

The spectrum of RSs for a symmetric triple well structure, also shown in \Fig{fig:fig1RSE}\,(a) (blue squares and red crosses), is quite similar to that of the double well. We see from the inset that in spite of a rather strong potential of the middle well perturbing the double well system, the wave numbers of the even parity RSs are only slightly modified, while those for the odd parity remain strictly the same, since $\Delta V(0)=0$. The only significant change observed in the spectrum is that there are two antibound states which appeared on the negative imaginary half axis. These antibound states are formed from the closest to the origin pair of normal RSs of the double well spectrum, as it was discussed in detail in~\cite{Tanimu}.

At the same time, the asymmetric triple well spectra are quite different, see Figs.~\ref{fig:figxRSE}\,(a) and \ref{fig:fig2RSE}. They also show the same quasi-periodicity with the period of about  $\pi/(2a)$, determined by the full width of the system $2a$, not changed by the perturbation. However, one can see additionally another quasi-periodic behavior of the RS wave numbers, with a larger period which depends on the position $b$ of the middle quantum well, as it is clear from \Fig{fig:fig2RSE}.

To study this effect, we have chosen the position of the middle well in such a way that it splits the system into two subsystems, with the smaller subsystem being $L$ times narrower than the full system. The results are shown in Figs.\,\ref{fig:figxRSE}\,(a) and \ref{fig:fig2RSE}\,(a-c) for $L=3$, 4, 5, and 10, respectively. The RS spectra for these systems are quasi-periodic, with $L$ neighboring RSs forming a period, as it is clear from these figures. Physically, this can be understood by looking again at the resonant condition for the constructive  interference of waves experiencing multiple reflections. Due to the commensurability of the widths of the full system and the smaller subsystem, the effect of constructive interference forming the RSs is enhanced for every $L$-th RS, owing to additional reflections from the middle well.

\subsection{Matrix elements of the perturbation}

To use the QM-RSE, one needs to calculate, using \Eq{Vnm}, the matrix elements of the perturbation \Eq{Vnm}. For the basis RSs wave functions given by
Eq.\,(\ref{SWE}) and the perturbation by \Eq{triplpot}, we find
\be
 \Delta V_{nm}=-\beta \varphi_n(b) \varphi_m(b)\,,
\label{pertmatx}
\ee
where
\be
\varphi_n(b)=2B_n\times\left\{
\begin{array}{cc}
      \cos (k_nb)  &  {\rm for\ even\ RSs}\,, \\
   \\
     i\sin (k_nb) &  {\rm for\ odd\ RSs}\,, \\
   \end{array}
\right.
\label{evodwave}
\ee
and the normalization constants $B_n$ are given by \Eq{norm-double}.

In general, if both the unperturbed and perturbed potentials are symmetric, the RS wave functions in each case  are either even or odd. In other words, the perturbation matrix $\Delta V_{nm}$ does not lead to any mixing of RSs of different parity. However, for the delta-like perturbation \Eq{pertmatx}, if it is symmetric, i.e. if $b=0$, not only even and odd states do not mix, but, moreover, odd basis RSs do not perturb. This is clear from the fact that the matrix elements are non-vanishing only between even parity states. In this case the matrix elements are given by
\be
\Delta V_{nm}=-4\beta B_nB_m\,.
\label{Vnm-even}
\ee
The vanishing effect of the perturbation on the odd RSs is also confirmed by the exact solution for the symmetric triple well presented in \Fig{fig:fig1RSE}\,(a) which shows that the wave numbers of the odd RSs of both double and triple quantum wells coincide.

%\newpage
\subsection{QM-RSE for a triple symmetric quantum well}
\label{b0}

We treat with the QM-RSE the symmetric triple quantum well first. The QM-RSE results are generated by solving numerically the linear matrix eigenvalue problem \Eq{RSE} with $\Delta V_{nm}$ defined by \Eq{Vnm-even}. The infinite matrix $H_{nm}$ in \Eq{RSE} is truncated in such a way that all RSs within a circle of radius $R$ centered at $k=0$ in the complex $k$-plane are kept in the basis. This introduces the total number $M$ of the basis RSs. We use this definition of the basis for all the examples treated in this paper.

We compare in \Fig{fig:fig1RSE}\,(a) the QM-RSE result for the RS wave numbers (red crosses) for a symmetric triple quantum well with the exact solution, Eqs.\,(\ref{triple}) and (\ref{xi1}) (blue squares). The unperturbed wave numbers for even and odd parity RSs are shown by black and green circles, respectively. We see that applying the perturbation does not change the wave numbers of the odd RSs of the basis system, as discussed above. At the same time, all even RSs are modified due to the perturbation, including the ground  state of the system (shown by the topmost square/cross on the imaginary $k$-axis).

It is clear from the comparison in \Fig{fig:fig1RSE}\,(a) that the QM-RSE is reproducing the exact values. 
The only RS having no exact solution to compare with is the antibound state with the wave number close to zero. The implemented procedure using the Newton-Rawson method failed to find the exact value in this case.
To quantify the agreement between the QM-RSE and exact values, we show in \Fig{fig:fig1RSE}\,(b) the relative error $|(\varkappa_\nu^{\rm RSE}-\varkappa_\nu^{\rm ex})/\varkappa_\nu^{\rm ex}|$, where $\varkappa_\nu^{\rm RSE}$ and $\varkappa_\nu^{\rm ex}$ are, respectively, the QM-RSE and the exact wave numbers of the perturbed RSs. The relative error is displayed for different basis sizes $M$ demonstrating the convergence of the QM-RSE to the exact solution as the basis size increases. Note that the shown values of $M$ include also odd basis states which remain unperturbed in this example. Figure~\ref{fig:fig1RSE}\,(b) allows us also to quantify the convergence: The relative error is approximately inversely proportional to the basis size $M$. Interestingly, for any fixed $M$ the relative error scales for different normal RSs as $1/k^2$, see the dashed line in \Fig{fig:fig1RSE}\,(b).

%This convergence is  rather slow compared to a similar problem in optics demonstrating a convergence proportional to $1/M^3$, see~\cite{Egor}.

\subsection{QM-RSE for a triple asymmetric quantum well}
\label{asym}

Applying the QM-RSE to an asymmetric triple quantum well structure shows a very similar accuracy and convergence of calculation, even though the perturbation now mixes even and odd
RSs of the basis system, effectively doubling the actual linear size of the matrix eigenvalue problem. Indeed, Figs.\,\ref{fig:figxRSE}\,(a) and \ref{fig:fig2RSE}\,(a-c) all demonstrate a visual agreement between the QM-RSE  and the exact solution, confirming the presence of the spectral changes, which are caused by the additional quantum interference effects in these structure, as discussed in \Sec{ansol}. Note that the Newton-Rawson solutions of the secular equations are much sensitive to the initial guess values used for finding the roots. As a consequence, some RSs can be missing in the analytic spectrum (which is not the case of the present calculation, but took place e.g. for the symmetric triple well problem, see \Fig{fig:fig1RSE}\,(a)). At the same time, the RSE finds {\it all solutions} in a selected spectral range and produces {\it no spurious solutions}. The reason for this is that the RSE is based on a complete (though, truncated) set of RSs of the unperturbed system used as an input, and as a result of the calculation, it returns as output a set of perturbed RSs which is also complete. Therefore, there can be no RSE solutions which are missing or spurious.

Figure \ref{fig:figxRSE}\,(b) demonstrates the convergence of the QM-RSE to the exact solution for the asymmetric triple well, which is very similar to the symmetric case.
The comparison with the exact solutions in Figs.\,\ref{fig:fig1RSE}-\ref{fig:fig2RSE} and the study of the relative errors for the RS wave numbers thus provides a verification of the QM-RSE in 1D. We can now take the advantage of the QM-RSE, applying it to more complex potentials, such as multiple quantum wells and finite quantum lattices, where the exact solutions are more difficult to find by other means.

\section{QM-RSE applied to finite quantum lattices}
\label{lattices}

In this section,  we apply the QM-RSE to finite periodic quantum potentials. We keep using the model of delta functions and define a finite periodic potential in such a way that it consists of $N$ equally spaced delta-like quantum wells of strength $\gamma$. The separation between the quantum wells, or the period of the potential is 
\be
d=\frac{2a}{N-1}\,,
\label{d-def}
\ee 
where $2a$ is the full width of the system, as before. The total potential of the system, $V(x)+\Delta V(x)$, thus consists of the unperturbed potential $V(x)$ of a double well, which is given by \Eq{unppot} and a perturbation
\be
\Delta V(x)=-\gamma\sum_{k=2}^{N-1}\delta(x-b_k)\,,
\label{PRDptbtn}
\ee
in which
\be
b_k=-a+d(k-1)
\label{position}
\ee
are the positions of the quantum wells. According to \Eq{Vnm}, the perturbation matrix of the QM-RSE is then given by
\be
\Delta V_{nm}=-\gamma\sum_{k=2}^{N-1}\varphi_n(b_k)\varphi_m(b_k)
\label{Pertmx}
\ee
with $\varphi_n(b)$ provided in \Eq{evodwave}.
\begin{figure}[t]
    \centering
\vskip0.3cm
\hskip-2.6cm
 \includegraphics[scale=0.35,angle=-90]{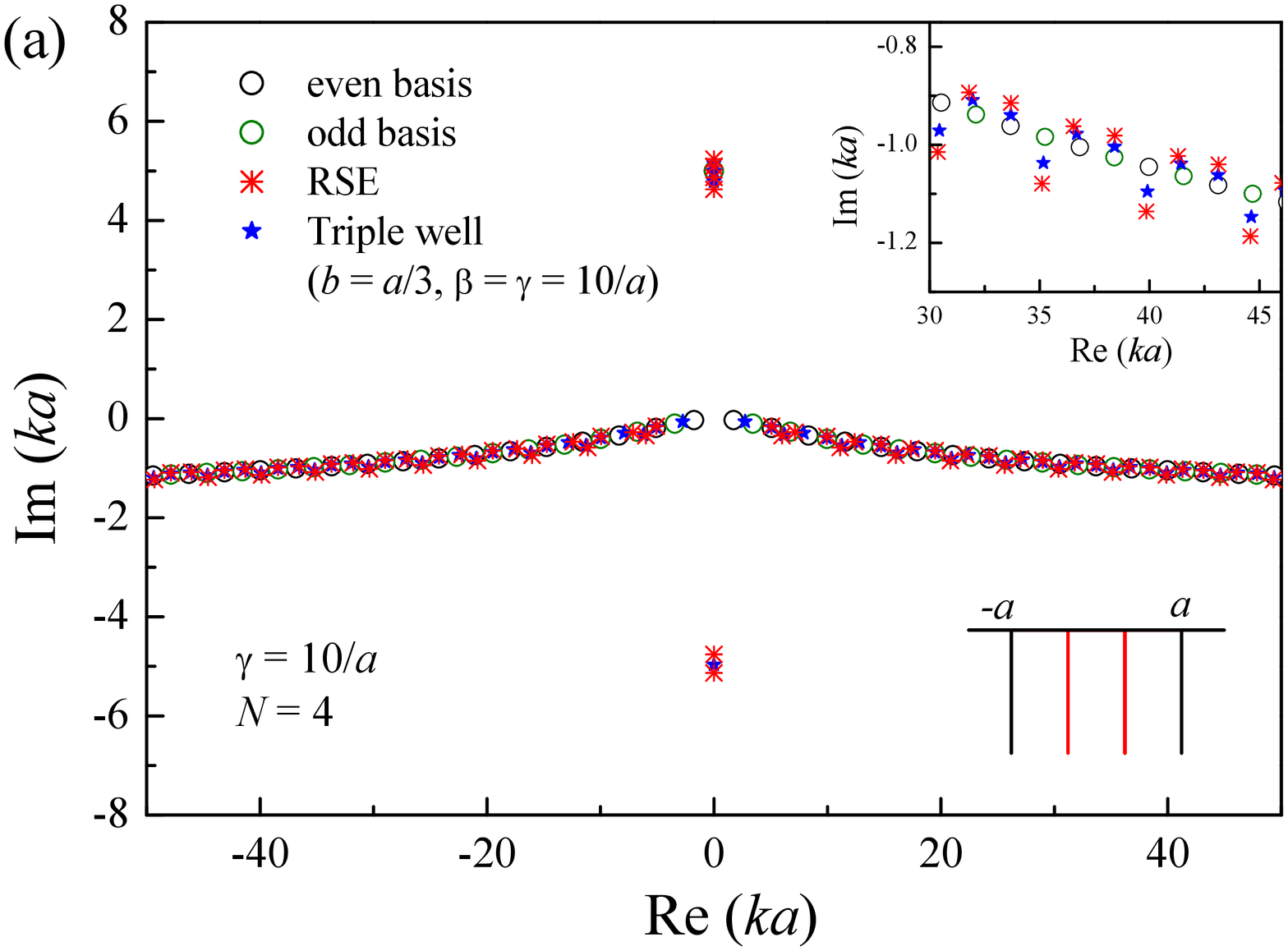}
\vskip-2.0cm
\hskip-2.6cm
 \includegraphics[scale=0.35,angle=-90]{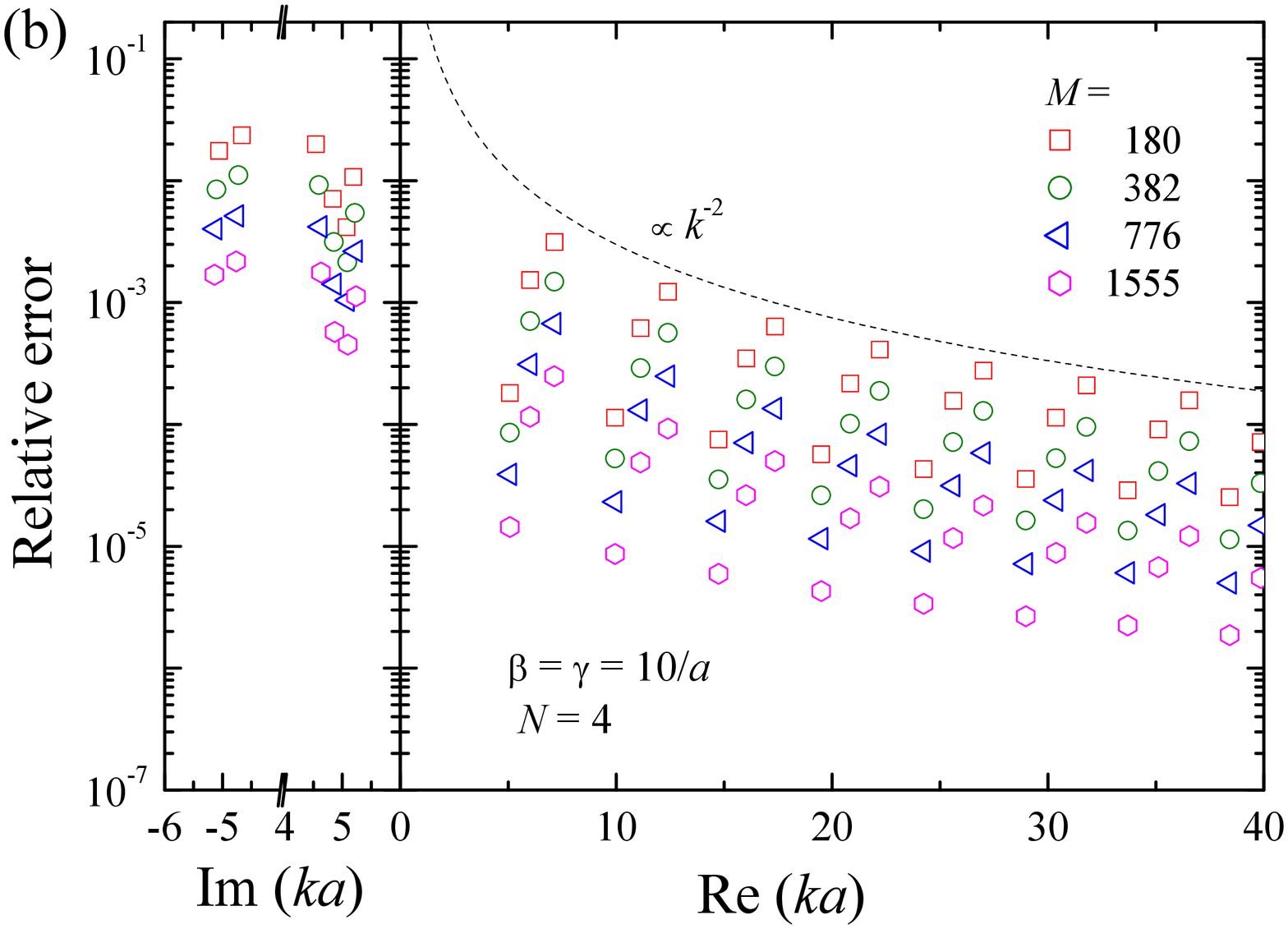}
\vskip-2.3cm
    \caption{As \Fig{fig:fig1RSE} but for a finite periodic potential with $N=4$ quantum wells of depth $\gamma=10/a$. The perturbation used in the QM-RSE is given by \Eq{PRDptbtn}. The resonant states of a triple well structure with  $\beta=\gamma=10/a$ and $b=a/2$ are shown additionally (blue stars).
    The relative error in (b) is calculated using the QM-RSE values for $M=4480$ replacing  the exact solution.
         }
\label{fig:errN5}
\end{figure}

We use the QM-RSE to calculate the RS wave numbers for increasing number of wells $N$. The $N=2$ case is the unperturbed system, and the $N=3$ case is already treated in \Sec{b0} above, see \Fig{fig:fig1RSE}. Therefore, we first look at the $N=4$ case. The wave numbers of both unperturbed and perturbed RSs for this case are shown in \Fig{fig:errN5}\,(a). The spectrum looks very similar to the ones considered before in what concerns the normal RSs, showing again a bigger period which we discuss below in more depth. However, a significant difference compared to the spectra in Figs.\,\ref{fig:figxRSE} and \ref{fig:fig2RSE} is the presence of two antibound states on the negative imaginary half-axis. 
\begin{figure}[t]
    \centering
\vskip0.3cm
\hskip-2.6cm
 \includegraphics[scale=0.35,angle=-90]{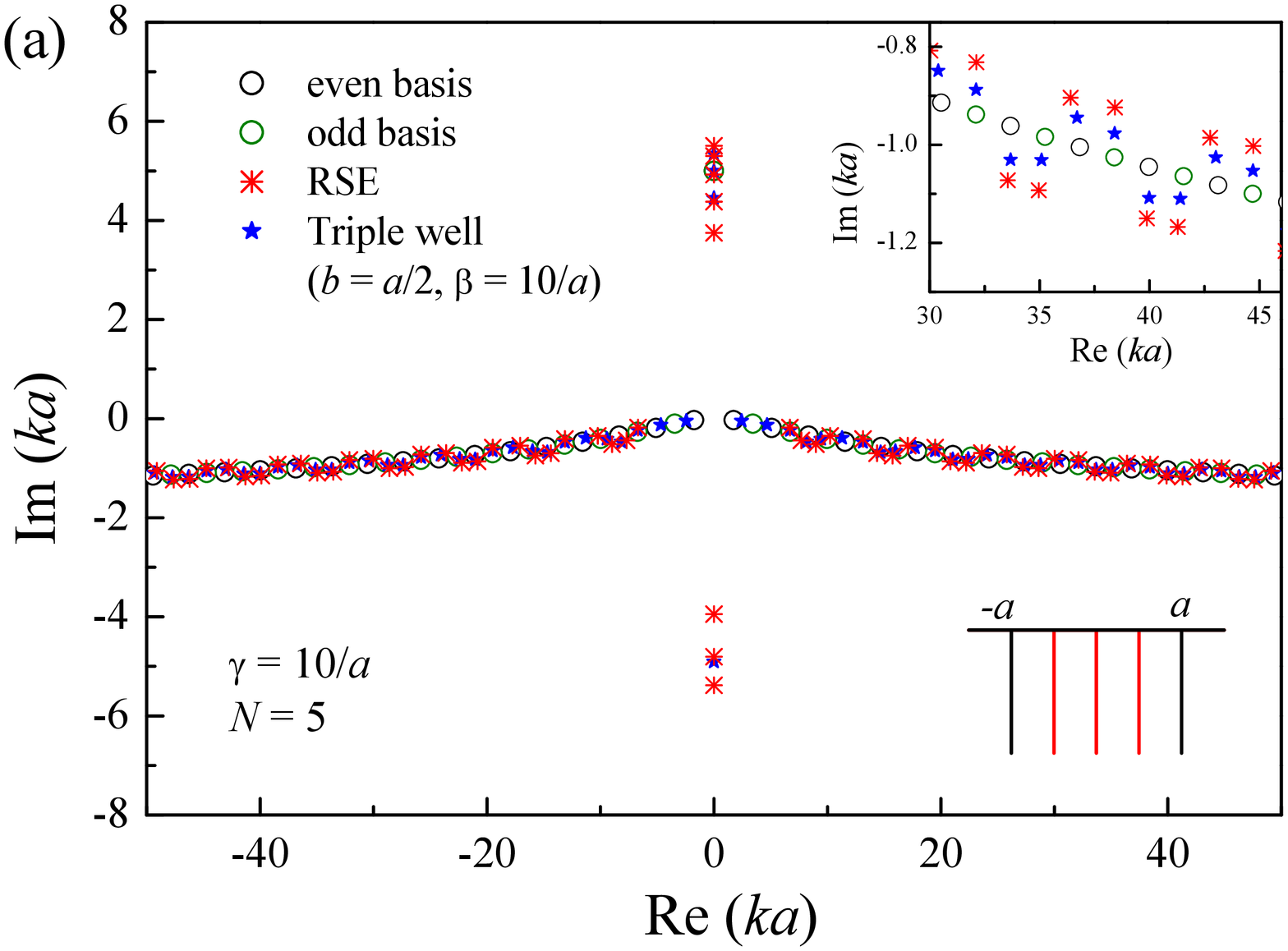}
\vskip-2.0cm
\hskip-2.6cm
 \includegraphics[scale=0.35,angle=-90]{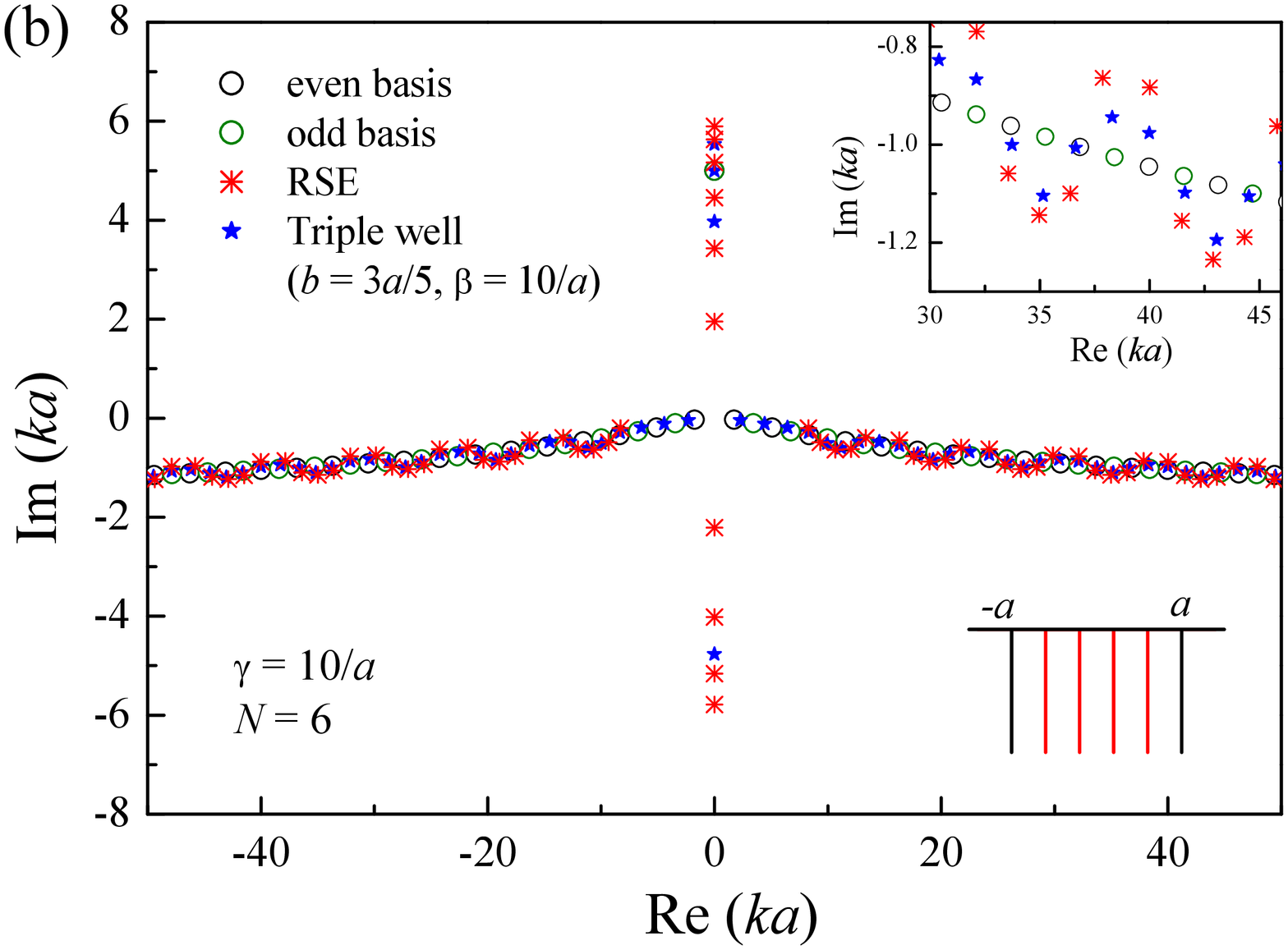}
\vskip-2.0cm
\hskip-2.6cm
 \includegraphics[scale=0.35,angle=-90]{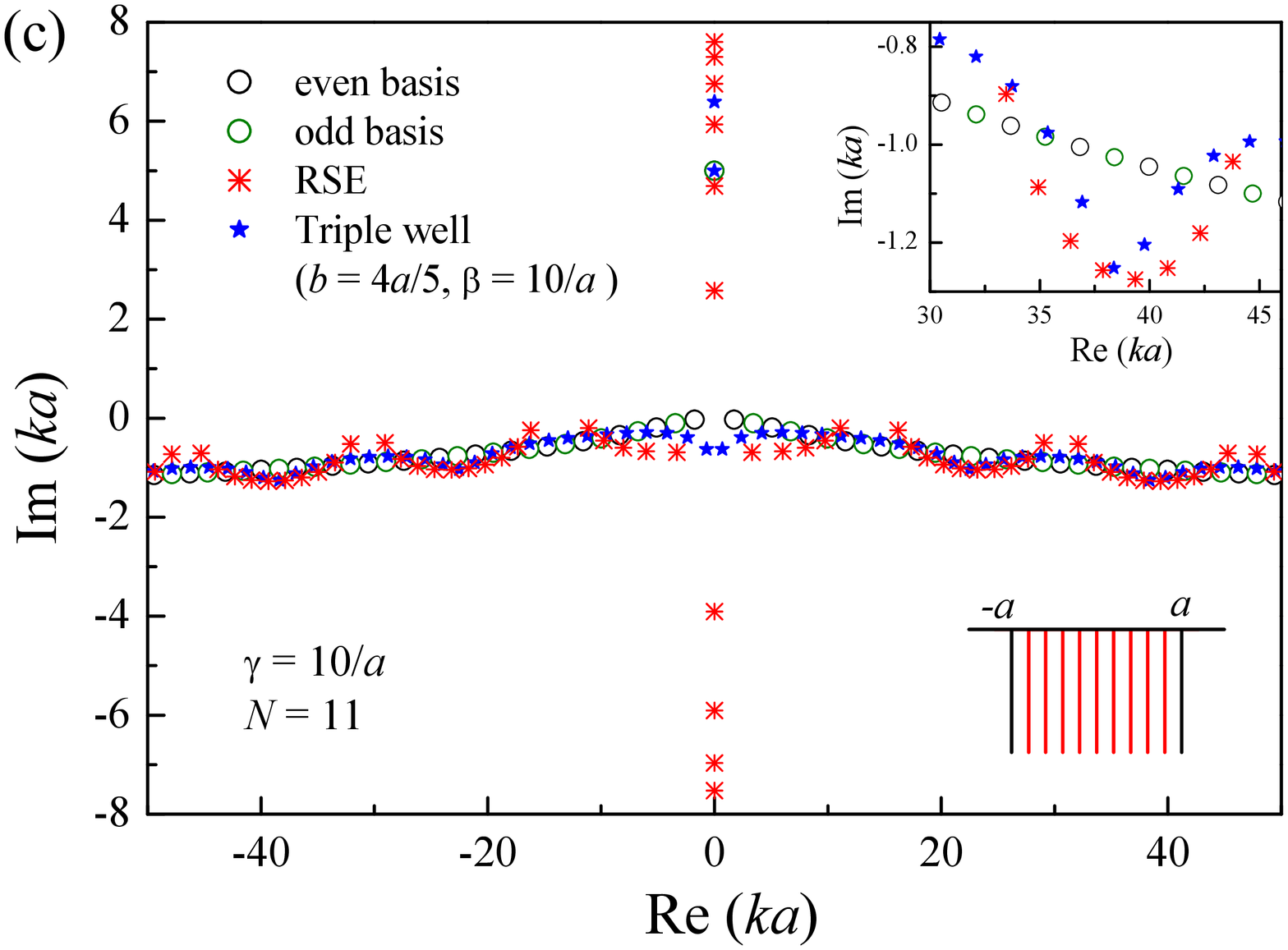}
\vskip-2.3cm
    \caption{As \Fig{fig:errN5}\,(a) but for $N=5$, 6, and 11. Additionally, the resonant states for a triple well structure with  $\beta=\gamma=10/a$ and $b=a/3$, $3a/5$, and $4a/5$ are shown, in (a), (b), and (c), respectively.}
\label{fig:N45611}
\end{figure}

This is due to two factors. First of all, the considered systems has a larger depth of the quantum wells: $\gamma=10/a$. Secondly, the total number of quantum wells is increased. Both factors result in a stronger overall quantum potential capable of accommodating a larger number of bound states: One can see that there are 4 bound states in this system, which are produced by a hybridization of the bound states of four individual quantum wells. Increasing the depth of the potentials reduces the tunnel coupling between the wells, which allows one to consider this coupling as a rather small perturbation, not affecting much the energy levels and keeping the number of states the same as without coupling. At the same time, the presence of two more bound states inevitably leads to two antibound states showed up in the spectrum. One can understand the presence of these two bound and two antibound states in the spectrum as a result of transformation of two pairs of normal RSs into bound and antibound states as the strength of the quantum wells increases, see~\cite{Tanimu} for a more detailed discussion of this phenomenon.

The convergence of the QM-RSE is quantified in \Fig{fig:errN5}\,(b) where we again show the relative error of the calculation of the RS wave numbers for four different basis sizes: $M\approx M_0$, 2$M_0$, 4$M_0$, and 8$M_0$. However, this time we do not find the exact solution, which would be a complicated, though not impossible task. Instead of the exact solution we take the values calculated with a much larger value of $M$. We see that the relative error is in principle very similar to that presented in Figs.\,\ref{fig:fig1RSE}\,(b) and \ref{fig:fig2RSE}\,(b), where the exact solution was used. Again, the $1/k^2$ dependence of the relative error for a fixed $M$ is observed, and the error scales inversely proportion to $M$.  We therefore conclude that the convergence of the QM-RSE does change when one makes the perturbation more complex.

\subsection{Comparison with triple well spectra}
Increasing $N$ further, the computational complexity of finding RSs using some alternative methods, such as transfer or scattering matrix approaches, increases dramatically. This is not only due to an increasing number of interfaces (or inhomogeneities) present in the system, determining the size of the linear algebra problem, but mainly because some of the eigenmodes are becoming prohibitively difficult to find, even when using the initial guess values in the Newton-Rawson method very close to the exact solution. These are usually the modes having the most interesting properties, such as superradiant states~\cite{Ivchenko} or high quality modes, which are similar in nature to bound states in the continuum~\cite{bound}. At the same time, the reliability of the QM-RSE remains the same, as well as the numerical complexity, provided that the integral strength of the perturbation did not change much. Indeed, in order to keep the accuracy of calculation the same, one needs to increase the basis size $M$ when the integral strength increases, which in turns affects the computational complexity scaling as $M^3$ owing to the matrix diagonalization required by the RSE.

Looking at \Fig{fig:N45611} where the RSs for $N=5$, 6 and 11 are shown, we see that the number of bound and antibound states increases with $N$ further (up to 6 and 4, respectively). However $N=6$ and $N=11$ lattices have the same number of bound/antibound states. This can be understood in the following way. The quantum tunneling between the wells increases with $N$, since the well separation $d$ decreases. Increasing the tunneling, one goes further away from the picture of nearly independent quantum wells (for which the tunneling is only a small perturbation, in which case the total number of bound states is equal to the number of wells). With increased tunneling, instead, the whole potential has to be treated more like one common wide quantum well which can accommodate a limited number of bound states. As for the antibound states, their number is usually two less the number of bound states~\cite{Tanimu}.

We also see in \Fig{fig:N45611} a quasi-periodic behavior of the RS wave number, similar to the phenomenon observed for triple quantum wells which is discussed in \Sec{ansol}.
Increasing $N$, the number of states in the period increases -- it is actually equal to $N-1$, as can be seen from the graphs. To confirm that this is a manifestation of the same effect (of an additional resonant enhancement, owing to the splitting of the whole system into two or more resonators), we compare in \Fig{fig:N45611} the RSs of finite periodic systems with those of the corresponding triple well system. We have chosen the smallest separation between the wells in the triple well system equal to $d$, the period of the quantum lattice, which is given by \Eq{d-def}. This comparison reveals close similarities between quantum lattices and the corresponding triple well systems. In particular, the quasi-periodic features observed in the spectra of both systems are almost the same. One can see from the top insets in Figs.\,\ref{fig:errN5}(a) and \ref{fig:N45611}(a-c) that the spectra of the two systems are in a good qualitative agreement. In other words, the RS spectrum of a finite periodic system does not change much if one removes from the potential all the inner quantum wells except the rightmost one. The physical reason for the quasi-periodic oscillations is essentially the same as mentioned above and discussed in more detail in \Sec{ansol}. The finite periodic potential with $N$ quantum wells splits the full system into $N-1$ resonators of width $d$ similar to the one such resonator present in the corresponding triple well. The presence of multiple resonators of a commensurable width in finite periodic structures only enhances the effect already observed in the triple wells: Indeed, the amplitudes of the quasi-periodic oscillations in the RS spectra are stronger in the case of the quantum lattices.

\subsection{Comparison with the Kronig-Penney model}
\label{sec-KP}
\begin{figure}[t]\centering
\vskip0.3cm
\hskip-2.6cm
 \includegraphics[scale=0.35,angle=-90]{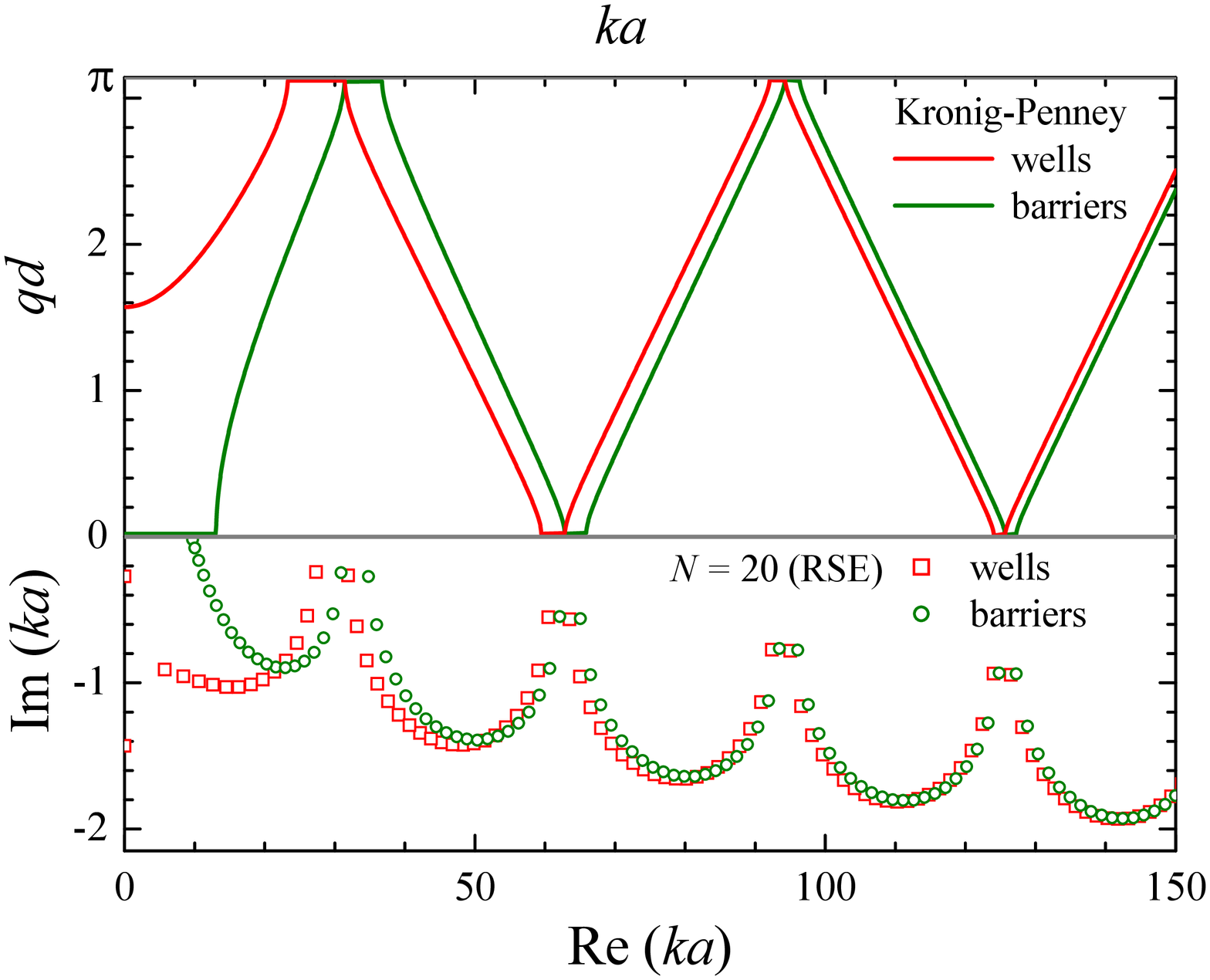}
\vskip-2.3cm
\caption{(top) Solution of the Kronig-Penney model  \Eq{KP9}. (bottom) Resonant state wave numbers of a finite periodic quantum lattice with $N=20$ calculated using the QM-RSE, for $\gamma=10/a$ (wells) and $\gamma=-10/a$ (barriers).
}
\label{fig:RSEKP}
\end{figure}

Taking the limit $N\to\infty$ while keeping the period $d$ (the distance between the neighboring wells) fixed, we end up with the famous Kronig-Penney potential~\cite{KP}
\be
U(x)=-\gamma\sum_{n=-\infty}^{\infty}\delta(x-nd)
\label{KP1}
\ee
describing an infinite periodic system, or an infinite quantum lattice. The Kronig-Penney model is known to have an exact analytic solution showing allowed bands and band gaps in the energy spectrum or the wave number spectrum of a particle. This exact solution is given by~\cite{KP}
\be
\cos(qd)=\cos(kd)-\frac{\gamma}{2k}\sin(k d)\,,
\label{KP9}
\ee
where $q$ is the wave number of the quasi-particle in the periodic potential, which is a conserved quantity. Indeed, according to Bloch's theorem, the wave function of the particle satisfies the periodic condition $\psi(x+d)=e^{iqd}\psi(x)$ which introduces this conserved wave number.

We compare in \Fig{fig:RSEKP} the spectra of RSs for a potential of $N=20$ wells (barriers) with the spectra of allowed and forbidden bands of a particle in the periodic potential \Eq{KP1}, corresponding to $N=\infty$. For the former, we again use the complex $k$-plane to display the RS wave numbers, see the bottom part of \Fig{fig:RSEKP}. For the latter, we use the $(q,k)$ plane with real values of $q$ and $k$, see the top part of \Fig{fig:RSEKP}. In this comparison, we use the same parameters of both structures: $\gamma=10/a$ and $d=2a/19$. 

One can see in \Fig{fig:RSEKP} a clear qualitative agreement between the allowed bands and the groups (periods) of RSs observed in the RS spectra of finite quantum lattices. The Kronig-Penney model thus helps us to clarifies on the actual physical meaning of these periodic groups of RSs: In the limit $N\to\infty$ they just form the allowed bands in the energy spectrum of the particle in a periodic potential. Figure~\ref{fig:RSEKP} shows results both for the wells and the barriers, demonstrating a good agreement and correlation between finite and infinite periodic structures.

\subsection{Varying the potential strength}
\begin{figure}[t]\centering
\vskip0.3cm
\hskip-2.6cm
 \includegraphics[scale=0.35,angle=-90]{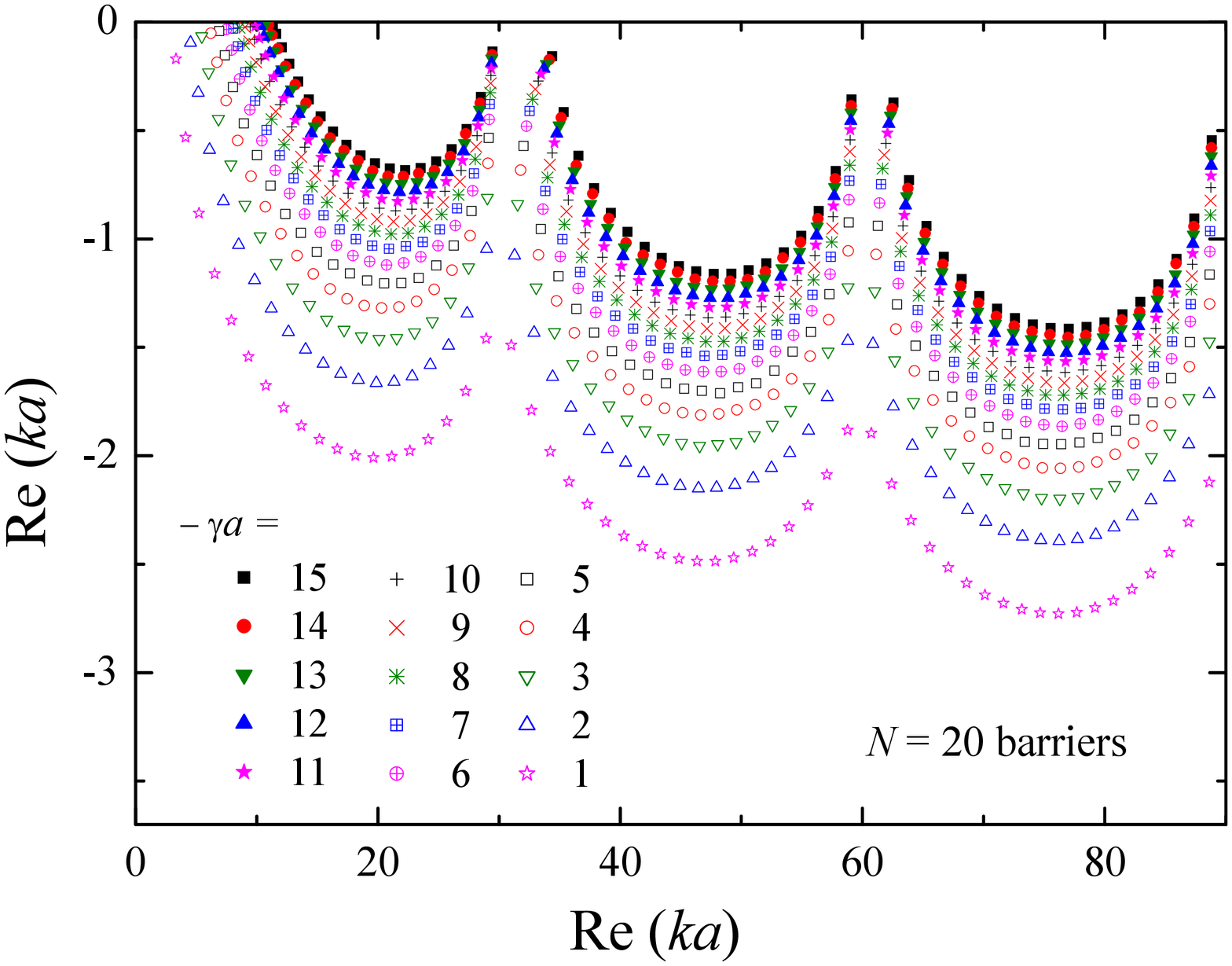}
\vskip-2.3cm
\caption{Resonant state wave numbers of a finite periodic lattice of $N=20$ quantum barriers, calculated using the QM-RSE for different barrier strength $\gamma$ as given.}
\label{fig:PERIODIC}
\end{figure}
We finally study the dependence of the RS wave numbers on the potential strength of a finite periodic structure of $N=20$ quantum barriers.
We see from \Fig{fig:PERIODIC} that each group of RSs of the quasi-periodic spectrum (discussed in \Sec{sec-KP}) is robust to varying the potential strength. However, the separations between the groups, which would correspond to the band gaps in the spectra of the ideal periodic system, strongly depends on $\gamma$. This is consistent with the result of the Kronig-Penney model, also showing a similar dependence of the band gap width on the potential strength. We do not provide here any quantitative comparison, though.

Another important effect is a decrease of the imaginary part of the RS wave numbers as the potential strength $\gamma$ increases. In other words, increasing $\gamma$ improves the quality factor (Q-factor) $Q=|{\rm Re} k_n/(2\,{\rm Im} k_n)|$ of all the resonances. This is expected, as higher values of $\gamma$ provide a better reflection from the potential inhomogeneities, in this way helping a certain probability density to stay longer within the system. For some resonant states, the Q-factor is becoming quite large, see, for example, the leftmost RS in \Fig{fig:PERIODIC} reaching $Q\approx 400$. The physical reason for the formation of such states could be similar to that of bound states in the continuum~\cite{bound}. The true bound states, however, would have an infinite Q-factor.

\section{Conclusion}
\label{concls}

We have applied the resonant state expansion (RSE), a novel powerful theoretical method recently developed in electrodynamics, to non-relativistic quantum-mechanical systems in one dimension, modeling all potentials with Dirac delta functions. We have verified the method, which we call here quantum-mechanical resonant state expansion (QM-RSE), testing it on systems with triple quantum wells while using the resonant states of a double quantum well as a basis. We have studied the convergence of the QM-RSE to the exact solutions. In particular, we have demonstrated that the QM-RSE is asymptotically exact, with the number of basis resonant states being the only technical parameter of the method, and that the relative error scales inversely proportional to the basis size. We have further demonstrated that the QM-RSE enables an accurate and efficient study of complicated quantum structures, such as multiple quantum wells and finite periodic potentials, which are harder to address by alternative methods. Some complicated quantum systems can exhibit interesting physical phenomena, such as formation of quasi-periodic bands of resonances or bound states in the continuum, and thus have to be investigated with an accurate and efficient tool.
The QM-RSE can offer such a tool, as we have demonstrated in this paper.

\end{document}